# Non-Hermitian acoustic metamaterial for the complete control of sound by accessing the exceptional points

Yi-Fan Zhu[1†], Xue-Feng Zhu[2†], Xu-Dong Fan[1], Bin Liang[1*], Xin-Ye Zou[1], Jing Yang[1], and Jian-Chun Cheng[1*]

[1]*Key Laboratory of Modern Acoustics, MOE, Institute of Acoustics, Department of Physics, Collaborative Innovation Center of Advanced Microstructures, Nanjing University, Nanjing 210093, P. R. China*

[2]*School of Physics, Huazhong University of Science and Technology, Wuhan, Hubei 430074, P. R. China*

[†]These authors contributed equally to this work.

[*]To whom correspondence should be addressed. Emails: jccheng@nju.edu.cn (J. C. C.), liangbin@nju.edu.cn (B. L.)

**Abstract**

Non-Hermitian systems always play a negative role in wave manipulations due to inherent non-conservation of energy as well as loss of information. Recently, however, there has been a paradigm shift on utilizing non-Hermitian systems to implement varied miraculous wave controlling. For example, parity-time symmetric media with well-designed loss and gain are presented to create a nontrivial effect of unidirectional diffraction, which is observed near the exceptional points (EPs) in the non-Hermitian systems. Here, we report the design and realization of non-Hermitian acoustic metamaterial (NHAM) and show that by judiciously tailoring the inherent loss, the phase and amplitude of reflection can possibly be tuned in a decoupled manner. Such decoupled tuning of phase and amplitude is closely related to the EPs. As a demonstration of functionality, we experimentally generate a high-quality acoustic hologram via NHAM. Our work may open a new degree of freedom for realizing the complete control of sound.

Loss is an inherent phenomenon accompanied with varied forms of energy. In acoustics, loss is usually caused by viscosity or impedance mismatch between two contrasted media. It is undesired for most cases in acoustic wave manipulation, owing to the fact that energy is non-conserved and the information carried by sound signals is lost. For example, acoustic metamaterial [1-24], as one kind of artificial materials with exotic properties, is normally treated as lossless or Hermitian [25]. However, in the overdamped resonance condition, inherent loss cannot be ignored and may eliminate the exotic properties of metamaterial. Recently, the field of non-Hermitian acoustics keeps on rising [26-30] and becomes a hot topic in material science, which shows the very possibility of making benefits from the engineered loss. The most representative example should be parity-time symmetric acoustic medium, where the judiciously designed loss and gain can lead to one-way reflection as well as unidirectional cloaking [26-30]. Those intriguing phenomena are demonstrated to be closely related to the existence of exceptional points (EPs) in the non-Hermitian systems.

In this paper, we harness the inherent loss in acoustic metamaterial to explore the new degree of freedom for the complete control of sound. The lossy metamaterial is an open system and thus termed non-Hermitian acoustic metamaterial (NHAM). Our proposed NHAM is a partially reflective holey structure with an open aperture perforated at the back side of each deep-subwavelength unit-cell. The open aperture brings controllable leakage of energy (or cavity loss) in each unit-cell. By engineering the loss judiciously from the aperture size in each unit-cell, the reflection amplitude can be modulated within a full range from 0 to 1 at will. On the other hand, the reflection phase is mainly determined by another geometric parameter of NHAM. As a result, we can possibly realize full controls of both amplitude and phase, making the current NHAM in stark contrast to the conventional phase-modulated metamaterial, *e.g.*, acoustic metasurface [31-39]. We further show that the amplitude-phase modulation in NHAM is much more interesting than the one in a Hermitian system due to the existence of EPs, where the controls of amplitude and phase are completely decoupled. This intriguing feature in the vicinity of EPs enables ergodicity in amplitude-phase modulation space, thus leading to the complete manipulation of sound.

In the end, we experimentally demonstrate the complete control of sound via NHAM by projecting a high-quality acoustic hologram. We know that optical holograms have significantly advanced with the cutting-edge nanotechnology and already changed our daily life dramatically [40-44]. We also expect that there would be many potential applications of acoustic holograms in the fields, such as particle manipulations, ultrasonic therapy, and imaging, *etc*. To our best knowledge, holograms are usually realized solely with phase modulation [41-43], which unavoidably requires optimizing algorithms, such as computer-generated hologram methods [44], to obtain the targeted phase profile. In our proposal, NHAM provides great flexibility in design of acoustic holograms, since both amplitude and phase information can be readily recorded in the metamaterial. As a result, acoustic holograms with high-quality and fine resolution can be achieved by simply using the time-reversal technique without employing complex optimization processes. In some previous works, researchers have used active elements to realize acoustic holograms, which, however, requires to operate individual sources with non-uniform properties and large pixel density, let alone the high cost of the complicated and expensive electronic controlling systems [45]. Our designed NHAM is capable to present continuous and precise amplitude and phase profiles, since the size of "pixel", *viz.*, the cross-sectional dimension of a unit-cell, can be tailored down to deep-subwavelength scale. Our proposal would yield rapidly varying spatial modulations of both amplitude and phase, and therefore guarantee high-quality holograms with large throughput, fine resolution and low background noise.

**Results**

**Design of NHAM.** First we explain the design principle of NHAM and clarify the underlying mechanism of simultaneous control of amplitude and phase, which goes beyond the existing acoustic metasurfaces with only phase modulation [31-39]. Figure 1(a) shows the schematic of NHAM comprising an array of unit-cells ($39 \times 39$), with the predesigned geometries as displayed in Figs. 1(b) and 1(c). The NHAM is a partially reflective holey structure, with the ability to modulate both amplitude and phase of reflection at the surface under the illumination of sound on the front side, indicated by the red arrows in Fig. 1(b). For each unit-cell, an open aperture is perforated on the back side. The leakage of sound through the aperture introduces the effective loss in the metamaterial, indicated by the blue arrow. Each unit-cell comprises three components: upper channel ($C_1$), middle channel ($C_2$), and lower channel ($C_3$), as shown in Fig. 1(c).

In this study, the operating wavelength is $\lambda$, the period of unit-cells $D = \lambda / 4$

(*viz.*, the size of pixel), and the speed of sound in air $c_0 = 343 \mathrm{ms}^{-1}$. Therefore, when we choose an operating frequency $f = 17150 \mathrm{Hz}$, the structrual period $D$ is set to be $0.5 \mathrm{cm}$. Due to the subwavelength nature of $D$, the amplitude and phase of reflection are independent of the incidence direction. Figure 1(c) shows the cross-section of a unit-cell, where the width of both $C_1$ and $C_3$ is $d = 0.4 \mathrm{cm}$ and the width of $C_2$ ($w$) is a variable. The channel walls are rigid since the acoustic impedance of solids is much larger than that of air. The heights of $C_1$, $C_2$, and $C_3$ are $h_1$, $h_2$, and $h_3$, respectively. For the planar metamaterial slab, the structural parameters $h_2$ and $h(= h_1 + h_2 + h_3)$ are fixed, where $h_2 = 0.5 \mathrm{cm}$ and $h = 2 \mathrm{cm}$. The effective acoustic parameters of NHAM can thus be engineered by adjusting $w$ and $h_1$. Figure 1(d) presents the relation between the amplitude $(r_A)$ and structural parameters $(w, h_1)$, while Fig. 1(e) displays the phase response $(\varphi)$ as a function of those two parameters. The results clearly show that the amplitude $r_A$ can be modulated within the range from 0 to 1, and the phase shift $\varphi$ can be modulated within the range from 0 to $2\pi$. However, we point out that this is only the first step in accessing all the possible combinations of $(r_A, \varphi)$. In general, $r_A$ and $\varphi$ are related to both $w$ and $h_1$ in such holy structures [46]. Therefore, they can be expressed into functions $r_A = f_1(w, h_1)$ and $\varphi = f_2(w, h_1)$. An intuitive way to ensure the ergodicity in phase-amplitude space is to make $r_A$ and $\varphi$ respectively controlled by one parameter, *viz.* $r_A = f_1'(w)$ and $\varphi = f_2'(h_1)$, which, as will be revealed later, is enabled by accessing the EPs in the NHAM. In Fig. 1(f), we calculated the phase and amplitude responses versus $w$ for the case of $h_1 = 0.75 \mathrm{cm}$. The result shows that by changing the parameter $w$, the amplitude of reflection can be controlled to cover the range from ~0 to 1 without much affecting the phase of reflection. In a similar way, Fig. 1(g) gives the calculated phase and amplitude responses versus $h_1$ for the case of $w = 0.2\text{cm}$. The result shows that by changing the parameter $h_1$, the phase of reflection can be tuned within a full range from 0 to $2\pi$, while the amplitude of reflection is almost unchanged. Figures 1(f) and 1(g) display the independent relations of $r_A$ and $\varphi$ with $w$ and $h_1$, respectively. The complete decoupling is a typical singularity effect closely related to the existing EPs in non-Hermitian systems.

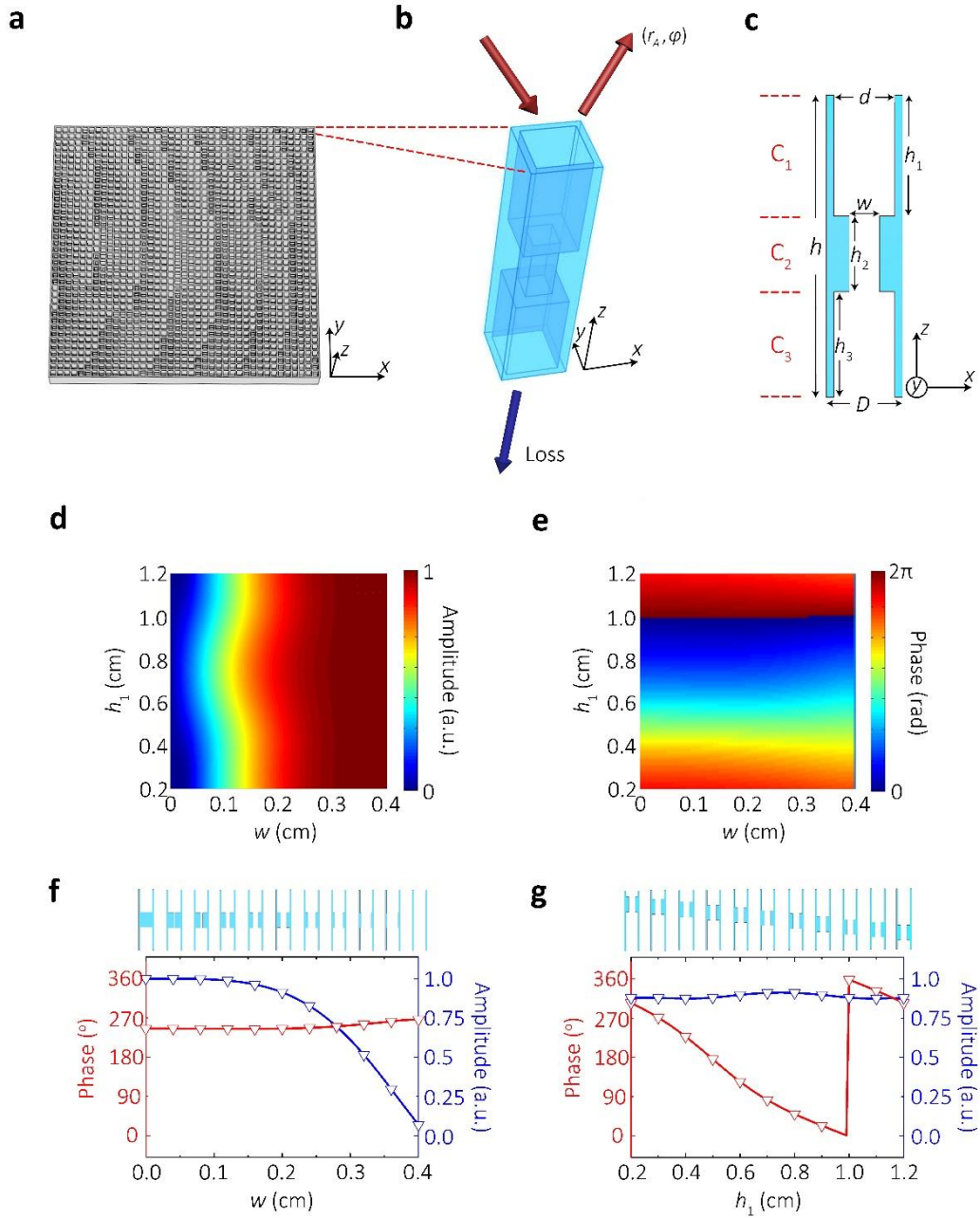

**Figure 1 | Design of NHAM.** (**a**) A schematic diagram of NHAM. (**b**) A 3D illustration of a unit-cell comprising three parts: $C_1$, $C_2$, and $C_3$. The incident waves are reflected at the front side of NHAM, indicated by the red arrows. The blue arrow shows the cavity loss at the back side. (**c**) A 2D illustration of a unit-cell where the structural parameters are marked. (**d**) and (**e**) The pseudo-color maps for the dependence of amplitude and phase of reflection ($r_A$ and $\varphi$) on the geometric parameters ($w$ and $h_1$). (**f**) The

amplitude and phase of reflection versus $w$ with $h_1 = 0.75 \mathrm{cm}$. (**g**) The amplitude and phase of reflection versus $h_1$ with $w = 0.2 \mathrm{cm}$.

**Complete decoupling at EPs.** To further demonstrate the complete decoupling at EPs, we define the coupling strengths $M_{rA}$ and $M_\varphi$ as

$$M_{rA} = \frac{\partial r_A}{\partial h_1} \frac{\partial r_A}{\partial w}, \tag{1}$$

$$M_\varphi = \frac{\partial \varphi}{\partial h_1} \frac{\partial \varphi}{\partial w}. \tag{2}$$

where $M_{rA}$ and $M_\varphi$ are the change rates for amplitude and phase of reflection with structural parameters, respectively. In Eqs. (1) and (2), we regard $M_{rA}(M_\varphi) = 0$ as the complete decoupling case, where the amplitude or phase of reflection is related to only one structural parameter ($h_1$ or $w$). We further obtain the coupling coefficients $\overline{M_{rA}}$ and $\overline{M_\varphi}$ by integrating the coupling strengths for all possible $(w, h_1)$ values and conducting normalizations with respect to their maxima. It is worth to be noted that the filling ratio of air channels, defined by $\beta = d / D$, is a very crucial factor to influence $\overline{M_{rA}}$ and $\overline{M_\varphi}$. For an extreme case of $\beta = 1$, *viz.* the channel walls are infinitely thin, the amplitude $r_A$ can be analytically derived as (See Supplementary Note 2 for details):

$$r_A = \frac{d^4 - w^4}{d^4 + w^4}. \tag{3}$$

Equation (3) clearly shows that the amplitude $r_A$ is only related to $w$, and we get $\overline{M_{rA}} = 0$ with a completely decoupling effect. The expression of $\varphi$ at $\beta = 1$ is (See Supplementary Note 2 for details):

$$\varphi = -\frac{4\pi h_1}{\lambda}, \qquad\qquad\qquad (4)$$

which is only related to $h_1$. Thus, we also have $\overline{M_\varphi} = 0$ at $\beta = 1$, showing a complete decoupling for the phase $\varphi$. The complete decoupling is a typical singular effect.

Therefore, the ideal condition $\overline{M_{rA}} = 0$ and $\overline{M_\varphi} = 0$ should correspond to the

singularity points (or EPs) in the non-Hermitian systems. In practice, we can access the

EPs by bringing $\beta$ close to 1, where the very large impedance contrast between solids and air plays an important role. However, the fact that the channel walls cannot be regarded as rigid when $\beta \to 1$ hinders us from exactly hitting the EPs. When $\beta < 1$,

the analytical expressions of $r_A$ and $\varphi$ are quite complicated due to the acoustic

impedance mismatches at the sudden-changed channel cross-sections (See

Supplementary Note 1). In this case, we can still achieve obvious quasi-decoupling

(coupling coefficients: ~0) effects by setting the parameter $h(= h_1 + h_2 + h_3)$ at specific values.

Figures 2(a) and 2(b) show the calculated coupling coefficients ($\overline{M_{rA}}$ and $\overline{M_\varphi}$)

for amplitude and phase of reflection with different structural parameters $(\beta, h)$. At

EPs where $\beta = 1$, as marked by the crosses in Figs. 2(a) and 2(b), the coupling

coefficients for amplitude and phase of reflection are both exactly 0. For the condition

of $\beta < 1$, the quasi-decoupling would occurs at some discrete $h$ values, *viz.*

$h = n\lambda / 2$ $(n = 2, 3, 4...)$. In this work, we set the filling ratio $\beta = 0.8$ and $h = 2\text{cm}$

to achieve the quasi-decoupling condition, as marked by the arrows in Figs. 2(a) and

2(b) (See Supplementary Note 3). In the vicinity of EPs, our designed metamaterial benefits from realizing all possible combinations of amplitude and phase of reflection by tuning two separate geometric parameters, *viz.* $h_1$ (the height of $C_1$) and $w$ (the width of $C_2$). As a consequence, the significant feature of independent phase and amplitude control enables the NHAM to control acoustic waves at will.

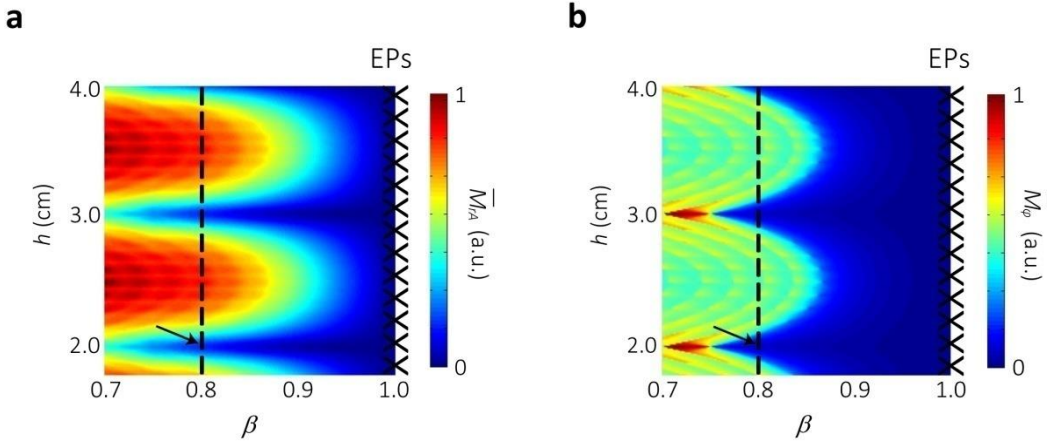

**Figure 2 | Coupling coefficient diagrams.** (**a**) The coupling coefficient $\overline{M_{rA}}$ for the amplitude of reflection with structural parameters $h$ and $\beta$. (**b**) The coupling coefficient $\overline{M_\varphi}$ for the phase of reflection with structural parameters $h$ and $\beta$. In (**a**) and (**b**), the EPs are located at $\beta = 1$, marked by the crosses. In this work, the quasi-decoupling is achieved at $\beta = 0.8$ with $h = 2\,\mathrm{cm}$, marked by the arrows.

**Design of an acoustic hologram.** As a proof of concept demonstration, we will generate a predesigned acoustic hologram with high quality by using NHAM. To this end, we first choose an image plane and set the virtual object to be projected in that plane. Here, we select three capital letters, *viz.*, "N", "J", and "U", as our virtual objects. By purposely designing the amplitude and phase profiles at the surface (or hologram plane) of NHAM, the incident wavefront can be reshaped to form the predesigned holographic pattern in reflection as shown in Fig. 3(a).

The positions of the hologram plane, the image plane, and the sound source are schematically identified in Fig. 3(b). The geometric size of our NHAM sample is $19.6 \times 19.6 \times 2 \text{cm}^3$ and the area of the image plane is around $20 \times 20 \text{cm}^2$. We set the image plane $22 \text{cm}$ away from the hologram plane. At the given frequency and incidence angle ($f = 17150 \text{Hz}$ and $\theta = 45°$), we record the phase and amplitude information into NHAM for projecting the holographic image. Without loss of generality, the NHAM is designed to give sound reflection normal to the hologram plane. It is worth mentioning that, our design approach is general and can be extended to generate high quality holographic images with much more complicated patterns in a preset image plane.

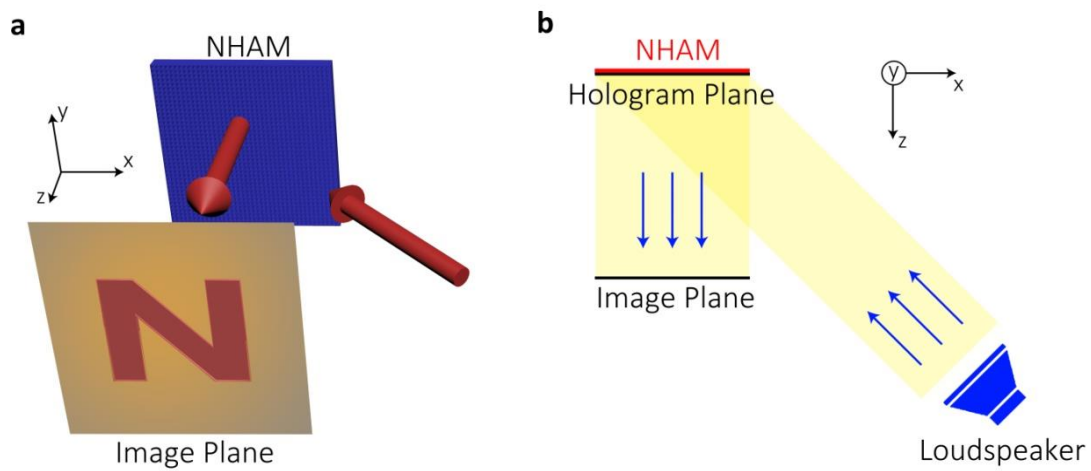

**Figure 3 | Schematic diagram of an acoustic hologram.** (**a**) A 3D schematic diagram of an acoustic hologram. A predesigned image (*e.g.*, the capital "N") can be generated by the reflection from NHAM at the given frequency and incidence angle. The arrows represent incidence and reflection directions of sound, respectively. (**b**) A schematic diagram showing the positions of the hologram plane, the image plane, and the sound

source.

Then we show how to produce the phase and amplitude profiles for projecting a predesigned image through the time-reversal technique. The predesigned image can be discretized into a collection of "pixels" with the size much smaller than the working wavelength. As shown in Fig. 4(a), the acoustic pressure $p(x, y)$ on the hologram plane can be calculated by superposing the wave components from all "pixels" of the predesigned image

$$p(x, y) = \sum_{j=1}^{N} \frac{A_j}{r_j} \exp(ik_0 r_j + \phi_j), \qquad (5)$$

where $N$ is the total number of "pixels" on the image plane, $A_j$ and $\phi_j$ are the amplitude and initial phase of the $j$th "pixel" at $(x_j, y_j, z_j)$, $r_j$ the distance between the $j$th "pixel" and the point at $(x, y, 0)$ on the hologram plane. Here we have

$$r_j = \sqrt{(x - x_j)^2 + (y - y_j)^2 + z_j^2}. \qquad (6)$$

Based on time-reversal symmetry, we can deduce the amplitude and phase profiles at the hologram plane by simply substituting Eq. (6) into Eq. (5). It should be pointed out that the conventional optic holograms of only phase modulation require a very complex optimization process for generating the targeted phase profile, where the amplitude is assumed to be uniform throughout the hologram plane [41-44]. In our case, we can break such limitation, thanks to the ability of independently modulating the amplitude and phase of reflection for all possible profiles. As a result, we can project high quality holographic images with a much simplified design process.

Figure 4(b) shows the pressure amplitude profile of a given hologram, which is chosen as the capital letter "N", comprising $91 \times 91$ "pixels" on the image plane. Figure 4(c) presents the calculated phase and amplitude profiles on the hologram plane for projecting the holographic image of the letter "N". The phase and amplitude profiles can be implemented by a well-designed NHAM sample as shown in Fig. 4(d). The NHAM sample has $39 \times 39$ unit-cells, which is fabricated by the 3D printing technology. The detail information of the NHAM sample is shown in the Methods section.

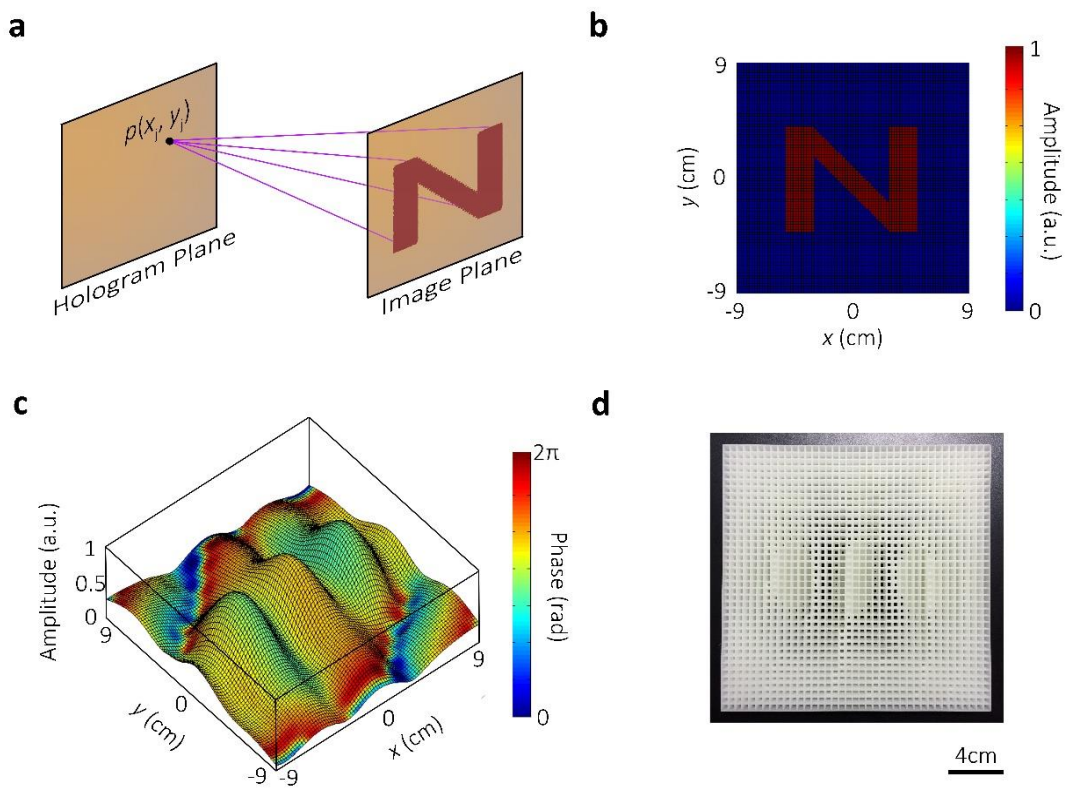

**Figure 4 | Design of an acoustic hologram.** (**a**) A schematic diagram showing the calculation of acoustic pressure $p(x, y)$ by superposing the wave components from all "pixels" of the predesigned image (the capital letter "N"). Thus we can generate the

desired image based on time-reversal symmetry. (**b**) The pressure amplitude profile of a given hologram (the letter "N"), comprising $91 \times 91$ "pixels" on the image plane. (**c**) The calculated phase and amplitude profiles on the hologram plane for generating a hologram of the letter "N". (**d**) A photograph of the NHAM sample for constructing the phase and amplitude profiles at the hologram plane.

**Experimental demonstration.** To make an experimental demonstration, we have fabricated three NHAM samples for projecting the predesigned holographic images of the capital letters "N", "J", and "U", respectively, at the frequency $f = 17150 \text{Hz}$. The pressure amplitude distribution is measured on the image plane (See the Methods section for experiment details). For comparison, we have also carried out analytical calculation and numerical simulation of the patterns on the image plane, as displayed in Figs. 5(a) and 5(b), respectively. Observation of Figs. 5(a)-5(c) verifies that the analytical and simulation results are in good agreement with the measurements, unequivocally showing that the three NHAM samples generate the high quality holographic images. We point out that the inevitable diffraction effect due to the finite size of unit-cells will lead to some discrepancy of the holographic image (Fig. 5) from the perfect one in Fig. 4(b). It would be possible to suppress this unwanted effect by increasing the unit-cell density in NHAM, since there is no cut-off for sound in deep-subwavelength channels.

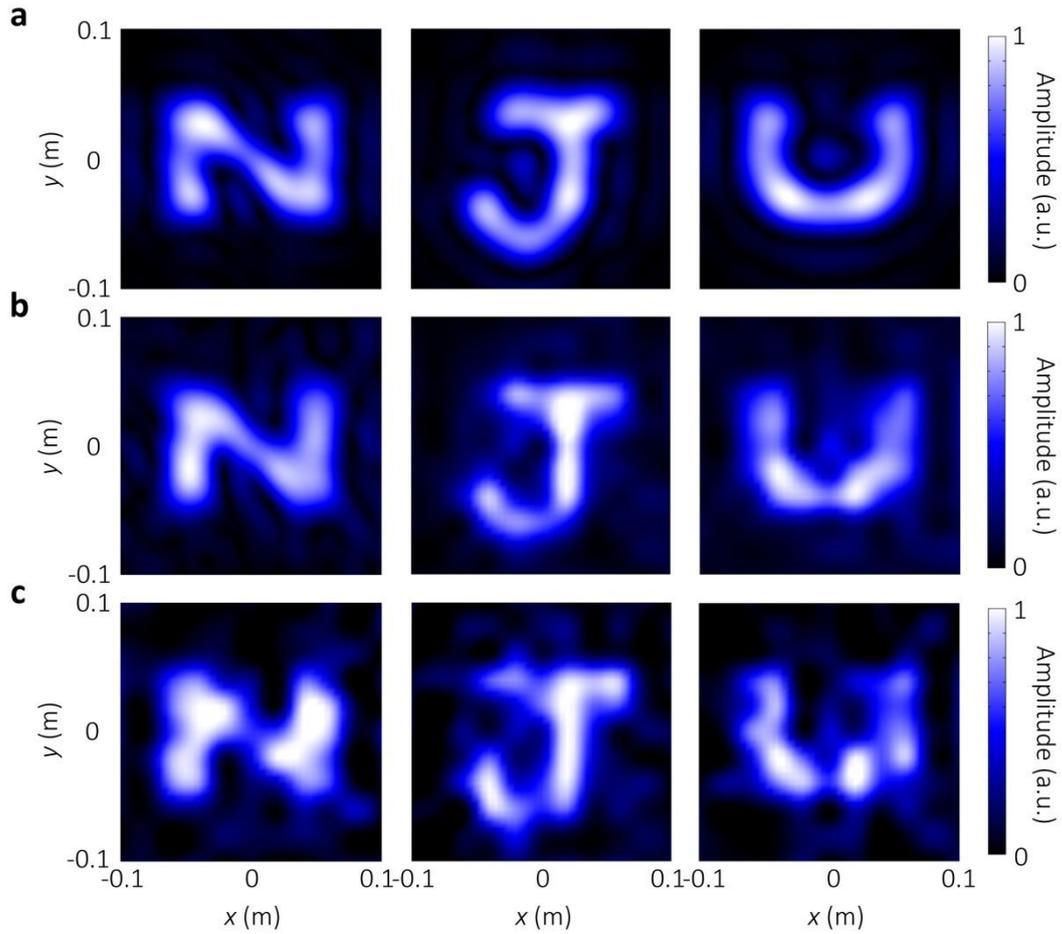

**Figure 5 | Experimental demonstration of acoustic holograms.** (**a**) Analytical results of holographic patterns (the capital letters "N", "J", and "U") on the image plane. The corresponding (**b**) simulation results and (**c**) experimental measurements of the holographic patterns on the image plane. The operating frequency $f = 17150\text{Hz}$.

## Discussion

We have proposed a new class of metamaterial with judiciously designed loss distribution and employed it to achieve the complete control of sound, *viz.*, modulating both amplitude and phase of sound in a precise, continuous and decoupled manner. To demonstrate the powerful wave manipulation ability of NHAM, we report the experimental demonstration of a high-quality acoustic hologram with large throughput, fine resolution, and low background noise. The independent control of phase and amplitude makes NHAM a very promising candidate for breaking the limitation of wave manipulation in the existing approaches. With NHAM, one can easily construct a required radiation pattern, regardless of the complexity, which is impossible by using the previously proposed metasurfaces and is expected to have fundamental significance in a great variety of important applications. For example, one can realize arbitrarily shaped focusing for HIFU treatments [47-49], which hitherto have to rely on the time-consuming scanning method, *etc.* We point out that NHAM also has the potential to realize complicated 3D holographic images by recording the targeted phase and amplitude profiles [41, 44], which is intriguing in particle manipulations [45], architectural acoustics [50], and so forth. We believe our findings will provide a capacious platform for the fundamental exploration of wave-manipulation physics as well as various acoustic-wave-based applications.

**Methods**

**Sample fabrication.** The NHAM samples are fabricated by 3-D printing machine (Stratasys Dimension Elite, 0.177mm in precision). The sample is fabricated by ABS (Acrylonitrile Butadiene Styrene) plastic of the density $\rho = 1180 \mathrm{kgm}^{-2}$ and the speed of sound $c = 2700 \mathrm{ms}^{-1}$.

**Experimental setup.** The experiment for acoustic holograms is performed in an anechoic chamber for reducing undesired sound reflection. In order to generate a plane wave, the sound source is located at far-field, about 3m away from the NHAM sample. The measured region is set at the image plane. The scanning of pressure amplitude field is conducted by using a 1/4-inch-diameter Brüel & Kjær type-4961 microphone and the software PULSE LABSHOP.

**Numerical simulation.** Three-dimensional numerical simulations are carried out by the finite element solver in commercial software COMSOL Multiphysics$^{\text{TM}}$ 5.0, using a high performance computing cluster. Perfectly matched layers are imposed on the outer boundaries of simulation domains to prevent reflections.

**Analytical calculation.** The pressure amplitude distribution in Fig. 5(a) is calculated by using $P(x, y, z) = \sum_{j=1}^{n} (r_{Aj} / r_j) \exp(ik_0 r_j + \varphi_j)$ , where $r_{Aj}$ and $\varphi_j$ are the amplitude and phase responses of the $j$th unit-cell of NHAM, $r_j$ the distance between the $j$th unit-cell of NHAM and the point at $(x_j, y_j, z_j)$ on the image plane, and $n$ the total number of unit-cells in NHAM. The coupling coefficients in Figs. 2(a) and 2(b) are calculated by using Eqs. (1) and (2), where the solutions of $r_{Aj}$ and $\varphi_j$ are shown in the Supplementary Note 1.

**Acknowledgements**

This work was supported by the National Basic Research Program of China (973 Program) (Grant Nos. 2011CB707900 and 2012CB921504), National Natural Science Foundation of China (Grant Nos. 11174138, 11174139, 11222442, 81127901, 11274168, and 11404125), NCET-12-0254, and A Project Funded by the Priority Academic Program Development of Jiangsu Higher Education Institutions.

**Author contributions**

Y.F.Z., X.D.F., X.Y.Z., and J.Y. performed the theoretical simulations; Y.F.Z. and X.D.F. designed and carried out the experiments; Y.F.Z., X.F.Z., B.L., and J.C.C. wrote the manuscript; X.F.Z., B.L., and J.C.C. guided the research. All authors contributed to data analysis and discussions.

**Additional information**

Competing financial interests: The authors declare no competing financial interests.